\documentclass[11pt,a4paper]{article}

\usepackage[utf8]{inputenc}
\usepackage[T1]{fontenc}
\usepackage{lmodern}
\usepackage{cite}
\usepackage[a4paper,margin=1in]{geometry}
\usepackage{microtype}
\usepackage{amsmath,amssymb,mathtools}
\usepackage{physics}
\usepackage{graphicx}
\usepackage{subcaption}
\usepackage{booktabs}
\usepackage{multirow}
\usepackage{xcolor}
\usepackage{hyperref}
\usepackage{float}
\usepackage{placeins}
\usepackage{xspace}
\hypersetup{colorlinks=true,linkcolor=blue,citecolor=magenta,urlcolor=blue}

\usepackage{enumitem}
\setlist{itemsep=2pt,topsep=3pt}

\renewcommand{\S}{$\mathbb{S}$\xspace}
\renewcommand{\H}{$\mathbb{H}$\xspace}
\renewcommand{\d}{\mathrm{d}}
\newcommand{\mga}{\textsc{MadGraph5\_aMC@NLO}\xspace}
\newcommand{\py}{\textsc{Pythia8}\xspace}

\title{\textbf{MC@NLO event generation by reweighting unweighted Born events}}
\author{
Saad El Farkh${}^{(a)}$\footnote{E-mail: \texttt{saad.el.farkh@cern.ch}}
\;,
Rikkert Frederix${}^{(b)}$\footnote{E-mail: \texttt{rikkert.frederix@fysik.lu.se}}
\; and\;
Mohamed Gouighri${}^{(a)}$\footnote{E-mail: \texttt{mohamed.gouighri@cern.ch}}
\vspace{6pt}
\\
{\small\it (a) Faculty of Sciences, Ibn Tofail University, Kenitra, Morocco}\\
{\small\it (b) Department of Physics, Lund University, Box 118, 221 00 Lund, Sweden}
}

\begin{document}
\date{}
\maketitle

\begin{abstract}
  \noindent
  We propose a computational strategy for NLO+PS simulations in the
  MC@NLO framework that starts from Born-accurate (LO) events and
  reweights them to the full MC@NLO \S-event weight, while generating
  \H-events separately. We validate the approach on two representative
  LHC processes and compare to direct NLO event generation for both
  standard MC@NLO and MC@NLO-$\Delta$ matching. Employing large
  folding values in the radiative variables stabilizes the \S-event
  integral, reduces weight variance, and significantly lowers the
  fraction of negative weights compared to \S-event generation without
  folding. At fixed precision, this pipeline has comparable wall-clock
  times relative to standard \S-event generation and unweighting, with
  room for further optimisation.
\end{abstract}

\vspace{0.5em}

\section{Introduction}
Simulating particle collisions at next-to-leading order (NLO) accuracy
matched to parton showers (PS) is essential for precision predictions
at the Large Hadron Collider (LHC). Among the established frameworks
for NLO+PS matching~\cite{Frixione:2002ik,Frixione:2007vw,Alioli:2010xd,Frixione:2010wd,Hoeche:2011fd,Platzer:2011bc,Alwall:2014hca,Jadach:2015mza,Frederix:2020trv,Sherpa:2024mfk,Sarmah:2024hdk,Bellm:2025pcw}, the MC@NLO method separates the hard
contributions into \S-events, which live on Born-level kinematics, and
\H-events, which contain an additional final state
parton~\cite{Frixione:2002ik}. While MC@NLO ensures NLO accuracy, its practical
implementation can be computationally demanding, particularly when
generating large samples of unweighted events for multi-particle
processes.

In this work, we present an alternative method for NLO event generation
within the MC@NLO framework. Our approach begins with the generation
of Born-level (LO) events, which are subsequently reweighted to obtain
the full MC@NLO \S-event weight. This reweighting incorporates
virtual corrections, integrated subtraction terms,
and the Monte Carlo counterterm; the latter require
numerically integrating over the radiative phase space associated with
each Born configuration. The \H-events are
generated separately using standard techniques.

By decoupling \S-event generation from the full NLO machinery and
using pre-generated unweighted LO samples, we aim to reduce the
computational cost of producing NLO-accurate event
samples. Furthermore, we employ large folding values in the radiative
variables~\cite{Nason:2007vt,Frixione:2007nu,Alioli:2010xd,Frederix:2020trv}
during the \S-event weight computation. This technique smooths the
integrand, reduces weight variance, and leads to a lower fraction of
negative weights compared to direct NLO \S-event generation without
folding---this improves the effective sample
size and downstream analysis efficiency.

We validate our method across two benchmark processes,
demonstrating that it exactly reproduces the physics of standard MC@NLO
simulations. Our results suggest that reweighting LO events to NLO
\S-events provides an valid alternative to
traditional NLO event generation workflows.

This paper is organised as follows. In section 2 we describe the
methodology for reweighting Born-level events to full MC@NLO S-event
weights, including the workflow of the implementation. In section 3
we validate the approach on the representative processes
$pp \to t\bar{t}$ and $pp \to e^{+}e^{-}j$, comparing reweighted
samples to directly generated MC@NLO S-events at the level of total
rates and differential distributions. Our conclusions and an outlook
for future developments are presented in section 4.

\section{Reweighting Born-level events to \S-events}
In the MC@NLO matching procedure, the two types of
contributions---$\mathbb{S}$-events and $\mathbb{H}$-events---can be
schematically written as:
\begin{align}
\mathbb{S}\text{-events:} \quad &
\left[
B(\Phi_B) +
V(\Phi_B) +
\int K_{\mathrm{MC}}(\Phi_R)\, \mathrm{d}\Phi_r
\right]
\times \mathcal{F}_{\mathrm{MC}}^{(B)}
\\
\mathbb{H}\text{-events:} \quad &
\left[
R(\Phi_R) - K_{\mathrm{MC}}(\Phi_R)
\right]
\times \mathcal{F}_{\mathrm{MC}}^{(R)}
\,,
\end{align}
where $B(\Phi_B)$ and $R(\Phi_R)$ denote the Born and real-emission
matrix elements squared, respectively. The phase-space measures are
$\d\Phi_B$ for the Born configuration and $\d\Phi_r$
for the radiative variables, with the full real-emission phase space
given by $\d\Phi_R = \d\Phi_B\, \d\Phi_r$.
For simplicity, the dependence on Bjorken $x$-fractions in hadronic
collisions, the PDF contributions, and the flux factors, are
suppressed in this schematic notation.

The term $K_{\mathrm{MC}}(\Phi_R)$ is the Monte Carlo (MC)
counterterm, representing the $\mathcal{O}(\alpha_s)$ expansion of the
shower emission probability. The integral over this term in the
\S-events is performed numerically, typically by using a single random
point in $\Phi_r$ per event, but approaches with using more points,
also known as folding, are possible as well.  To regulate this
integral, a local subtraction, such as the one given by the FKS
method~\cite{Frixione:1995ms}, is understood.

The virtual term $V(\Phi_B)$ includes the finite part of the one-loop
correction, along with remnants of the integrated local subtraction
terms (as well as mass factorization contributions in the case of
hadronic initial states). Finally, $\mathcal{F}_{\mathrm{MC}}^{(X)}$
denotes the standard parton-shower generating functional, where $X \in
\{B, R\}$ indicates whether the shower evolution starts from a
Born-level or real-emission configuration, respectively.

In the MC@NLO implementation in \mga, both \S- and \H-events are
generated dynamically and together during the event generation
phase~\cite{Alwall:2014hca}. In contrast, the approach presented in this work
decouples the generation of \S-events from the full NLO machinery by
starting from a pre-generated sample of Born-level events. For each
Born phase-space point $\Phi_B$, the full \S-event weight is computed
by numerically integrating over the radiative phase space
$\Phi_r$. That is, the unweighted Born-level events become \S-events
with their weights given by
\begin{equation}
  \label{weights}
w(\Phi_B) =
1+\delta \qquad \textrm{with}\quad \delta=\frac{
V(\Phi_B) +
\int K_{\mathrm{MC}}(\Phi_R) \d\Phi_r
}{B(\Phi_B)} \, .
\end{equation}
Once the reweighted \S-events are obtained, they are passed to the
parton shower using the generating functional
$\mathcal{F}_{\mathrm{MC}}^{(B)}$, consistent with the shower profile
used in the definition of $K_{\mathrm{MC}}$. The resulting event sample
is weighted, but can, in principle, be re-unweighted (up to a sign) by
a secondary unweighting procedure.

Within this approach the \H-events are generated separately using the
standard MC@NLO machinery. This separation ensures that the full NLO
accuracy is retained while allowing for an alternative treatment of
the \S-event component.

\subsection{Weighted samples and effective statistics}
Since reweighting produces \emph{weighted} \S-events, the Kish
effective sample size~\cite{kish},
\begin{equation}
  N_{\mathrm{ESS}} = \frac{(\sum_i w_i)^2}{\sum_i w_i^2} \equiv f_{\mathrm{ESS}}N\, ,
\end{equation}
is smaller than the number of events $N$. Note that even for
\emph{unweighted} \S-events, i.e., all $|w_i|$ equal, $f_{\mathrm{ESS}}$ can be
significantly smaller than $1$ when there is a large fraction of
negatively weighted events, see e.g.~ref.~\cite{Frederix:2020trv}.

Since $\delta$ starts at $\mathcal{O}(\alpha_S)$, naively it can be
expected that the weights $w$ form a narrow distribution around
$1$. As we will show below this is \emph{not} the case if no folding
is applied to the integral over the MC subtraction term: locally
the (subtracted) $K_{\mathrm{MC}}(\Phi_R)$ integrand can be
sizeable. However, with folding the situation greatly improves, even
so far that the effective sample size of a reweighted sample
(generated with folding) can be significantly larger than an
unweighted sample of \S-events generated without folding, due to the
latter having a much larger fraction of negative weights.  Of course,
with the same level of folding, the ESS of the unweighted sample will
always be larger than the sample generated through reweighting. But
starting from an unweighted Born-level sample means that the
$V(\Phi_B) + \int K_{\mathrm{MC}}(\Phi_R) \d\Phi_r$ contribution needs
to be evaluated orders of magnitude fewer times than in the default
MC@NLO setup\footnote{Note that in the standard \mga implementation,
a surrogate for the virtual corrections based on the Born matrix elements
squared is dynamically generated, greatly reducing the number of times
the virtual matrix elements need to be evaluated~\cite{Alwall:2014hca}. In the
reweighting procedure we do not use this surrogate: rather we compute the
exact virtual corrections for every unweighted Born event that we
reweight.}, retaining a much greater computational budget for folding
to a much higher degree.

\subsection{Workflow and implementation}

The reweighting procedure has been implemented within the \mga
framework~\cite{Alwall:2014hca}.  The workflow proceeds as follows:

\begin{enumerate}
\item \textbf{Process setup:} A process including NLO QCD corrections
  is generated using the standard \texttt{generate <process> [QCD]}
  command.

\item \textbf{LO event generation:} Unweighted Les Houches events
  (LHE) are produced at Born-level accuracy by selecting the
  appropriate flag at launch time. For each event, the random numbers
  used in its generation are stored alongside the event record. In
  \mga, each integration channel generates a separate event file
  before the events are merged in the final LHE file in the last step.

\item \textbf{Reweighting step:} Before the events are merged into a
  single file, the reweighter code reads each Born event,
  reconstructs it from the stored random numbers, and computes the
  corresponding \S-event weight according to eq.~\eqref{weights}. The
  output includes the new weight and updated shower starting scale(s)
  and leading-colour assignment where required. Performing the
  reweighting prior to file merging ensures that the integration
  channel used for event generation is unambiguously identified, which
  simplifies reconstruction and minimizes modifications to the default
  code base.
  
\item \textbf{Combining event files:} When the reweighting is
  completed, the separate event files corresponding to the integration
  channels are combined into a single LHE file, that can be used as
  a seed for the parton shower as usual.
\end{enumerate}

Since the only new feature is the execution of the reweighter code
(together with the trivial addition of storing the random numbers
along each event), this design keeps changes to the existing \mga
infrastructure minimal. This, therefore, reduces the likelihood of
implementation errors and bugs.

The \H-event contribution, that complements the \S-event part, can be
conveniently generated within the same process folder, ensuring
consistent parameter definitions. In practice, it is necessary to
select event generation at NLO accuracy at launch time, and setting
the newly-introduced \texttt{only\_h\_events} parameter to
\texttt{True} in the \texttt{run\_card.dat}.

\section{Results}
We consider two representative processes to test the idea of
reweighting LO events to \S-events,
\begin{align}
  &pp \to t\bar{t} \qquad \textrm{and}\\
  &pp \to e^+e^-j,
\end{align}
using the default setup in \mga. We generate 100k unweighted events at
LO accuracy and reweight with five different folding levels to
\S-events. We also produce separate samples where we directly generate
100k unweighted (up to a sign) NLO-accurate \S-events with three
different levels of folding.  Matching to the \py parton
shower~\cite{Bierlich:2022pfr} is enabled, without turning on
hadronisation or MPI, considering both the default MC@NLO and the
MC@NLO-$\Delta$~\cite{Frederix:2020trv} algorithms.

\begin{table}[h]
\centering
\small
\begin{tabular}{llcccr}
\toprule
$pp\to t\bar{t}$ &&
$\sigma$ [pb] &
$\delta\sigma$ [pb] &
$f_{\mathrm{ESS}}$ & time [s] \\
\midrule
\multirow{4}{*}{MC@NLO}
&RW\_LO (1,1,1)       & 778.2 & 7.6 & 0.10 & 229 \\
&RW\_LO (2,2,1)       & 768.3 & 4.1 & 0.34 & 266 \\
&RW\_LO (4,4,1)       & 770.2 & 3.5 & 0.49 & 465 \\
&RW\_LO (8,8,1)       & 767.4 & 3.2 & 0.59 & 1186 \\
&RW\_LO (16,16,1)     & 766.7 & 3.0 & 0.65 & 4178 \\
&NLO \S-events (1,1,1)& 766.3 & 2.9 & 0.69 & 533 \\
&NLO \S-events (2,2,1)& 768.3 & 2.5 & 0.92 & 1161 \\
&NLO \S-events (4,4,1)& 769.4 & 2.5 & 0.96 & 3549 \\
\midrule
\multirow{4}{*}{MC@NLO-$\Delta$}
&RW\_LO (1,1,1)       & 621.9 & 4.2 & 0.21 & 323 \\
&RW\_LO (2,2,1)       & 623.3 & 3.0 & 0.43 & 430 \\
&RW\_LO (4,4,1)       & 627.6 & 2.5 & 0.62 & 938 \\
&RW\_LO (8,8,1)       & 625.1 & 2.3 & 0.72 & 2998 \\
&RW\_LO (16,16,1)     & 628.4 & 2.2 & 0.77 & 11471 \\
&NLO \S-events (1,1,1)& 627.2 & 2.4 & 0.69 & 1027 \\
&NLO \S-events (2,2,1)& 627.1 & 2.1 & 0.90 & 2212 \\
&NLO \S-events (4,4,1)& 625.7 & 2.0 & 0.94 & 6917 \\
\bottomrule
\end{tabular}
\caption{The \S-events cross section $\sigma$, statistical
  uncertainty $\delta\sigma$ for 100k events, and the Kish effective
  sample size fraction $f_{\mathrm{ESS}}$ for the $pp \to t\bar{t}$
  process, with MC@NLO (top rows) and MC@NLO-$\Delta$ (bottom rows)
  matching. The final column shows the generation time of the LHE file.}
\label{tab:stats_ttbar}
\end{table}

In table.~\ref{tab:stats_ttbar} we summarise the results for the
$pp\to t\bar{t}$ process. The reweighted LO samples, with different
levels of folding, denoted by ``RW\_LO ($X$,$Y$,$Z$)'', for the levels
of folding for the three FKS variables, $\xi_i$, $y_{ij}$ and
$\phi_i$, respectively. Considering the statistical uncertainty
$\delta\sigma$, the cross sections $\sigma$ for the three reweighted
samples agree with the directly produced NLO \S-events
samples. Without any folding, the effective sample size
($f_{\mathrm{ESS}}=0.10$ and $0.21$ for MC@NLO and MC@NLO-$\Delta$,
respectively) is much lower than for the directly produced NLO
\S-event sample, $f_{\mathrm{ESS}}=0.69$ for both MC@NLO and
MC@NLO-$\Delta$. This comes as no surprise: the NLO \S-events are
unweighted (up to a sign), while the reweighted samples do not have
unit weights. Interestingly, with a significant amount of folding for
the reweighted samples, the effective sample size becomes close (for
MC@NLO) or even supersedes (for MC@NLO-$\Delta$) the unweighted NLO
\S-event effective sample size without folding. However, with folding
turned on for the direct NLO predictions, the latter have an effective
sample size that always significantly exceeds the reweighted
samples. The time it takes to generate each LHE sample is given in the
final column, computed as if running on a single CPU core. It can be
concluded that for this process, reweighting a sample is not
competitive, in both effective sample size and computational time,
with directly generating events at NLO.

\begin{figure}[!h]
\centering
\includegraphics[width=0.9\linewidth]{"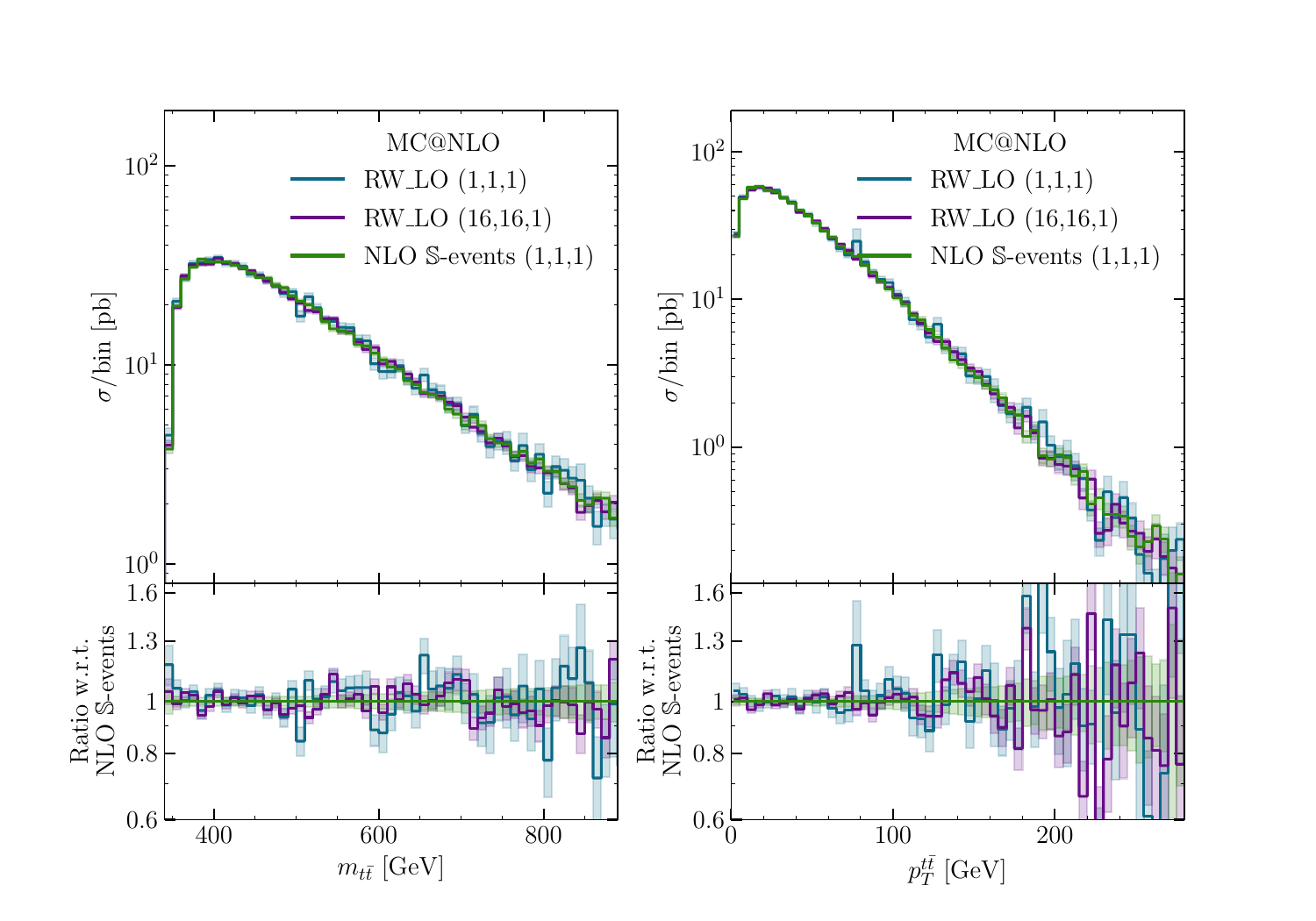"}
\includegraphics[width=0.9\linewidth]{"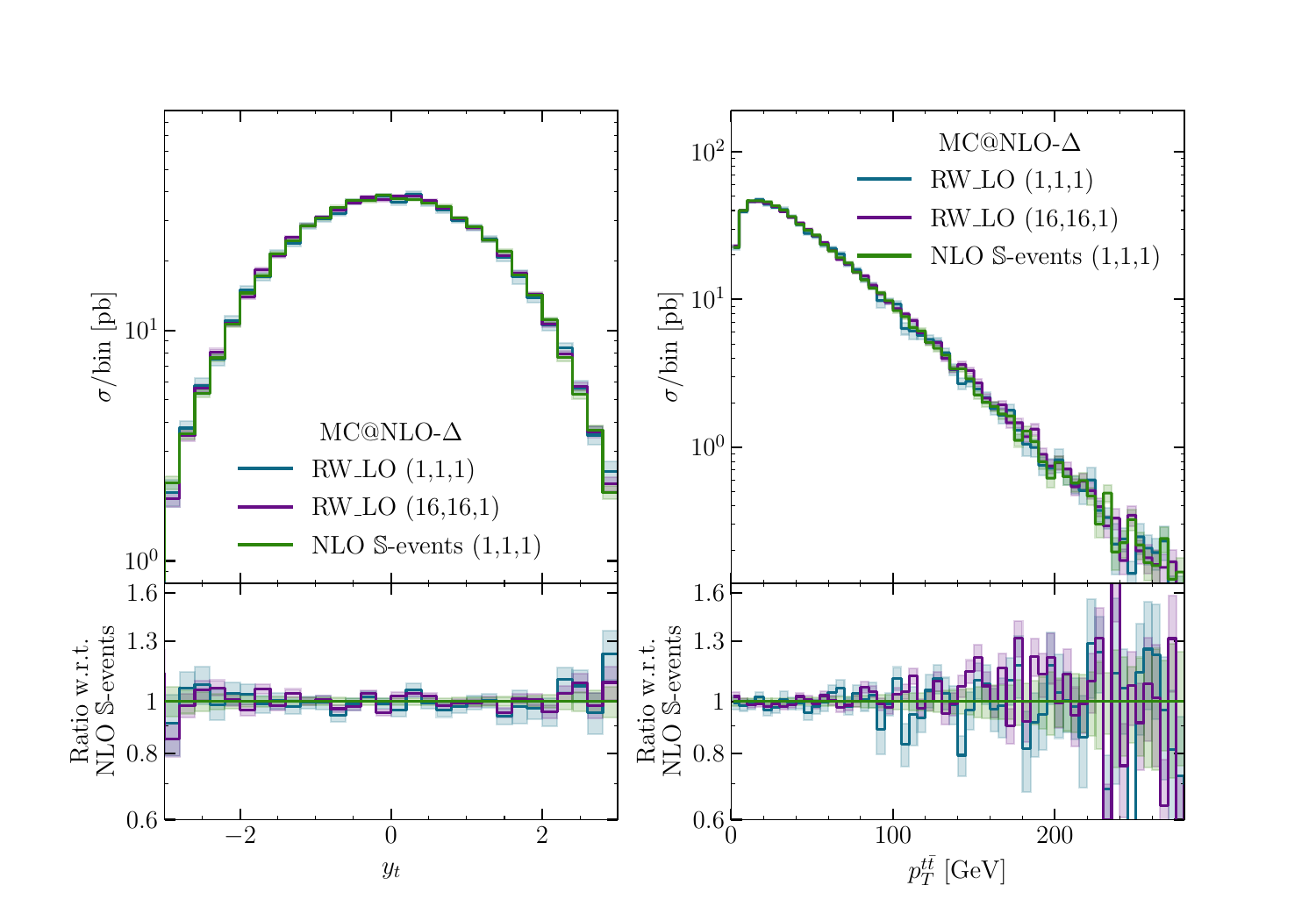"}
\caption{Differential distributions for the \S-event contribution to
  the process $pp \to t\bar{t}$ comparing reweighted LO samples and
  directly generated NLO S-events.  The top row shows results with
  standard MC@NLO matching: the $t\bar{t}$ invariant mass
  $m_{t\bar{t}}$ (left) and the transverse momentum of the $t\bar{t}$
  system $p_{T}^{t\bar{t}}$ (right).  The bottom row shows the
  corresponding results for MC@NLO-$\Delta$: the top-quark rapidity
  $y_t$ (left) and $p_{T}^{t\bar{t}}$ (right).  For each observable,
  we compare reweighted samples without folding and with high folding
  $(16,16,1)$ to directly generated NLO S-events without folding, with
  uncertainty bands indicating statistical errors.}
\label{fig:ttbar}
\end{figure}

In figure~\ref{fig:ttbar}, four representative plots are shown where
we compare a subset of the samples more differentially. The top two
plots are for MC@NLO matching, the bottom two for MC@NLO-$\Delta$. For
MC@NLO, we show the top--anti-top invariant mass, $m_{t\bar{t}}$,
(left) and transverse momentum of the $t\bar{t}$ system,
$p_{T}^{t\bar{t}}$, (right). For MC@NLO-$\Delta$, we show the rapidity
of the top quark $y_t$ is shown in the left plot, while in the right
plot we present the results for $p_{T}^{t\bar{t}}$. Similar
conclusions as we found for the total rates also apply here: the
results are in agreement within statistical uncertainties (denoted by the
semi-transparent bands), with the statistical fluctuations for the
reweighted sample without folding being significantly larger than for
the other samples.

\begin{table}[h]
\centering
\small
\begin{tabular}{llcccr}
\toprule
$pp\to e^+ e^-j$ &&
$\sigma$ [pb] &
$\delta\sigma$ [pb] &
$f_{\mathrm{ESS}}$ & time [s] \\
\midrule
\multirow{4}{*}{MC@NLO}
&RW\_LO (1,1,1)        & 1019.1 & 23.4 & 0.02 & 370 \\
&RW\_LO (2,2,1)        & 1003.7 & 11.8 & 0.07 & 441 \\
&RW\_LO (4,4,1)        &  999.3 &  7.8 & 0.16 & 727 \\
&RW\_LO (8,8,1)        & 1002.6 &  6.1 & 0.27 & 1750 \\
&RW\_LO (16,16,1)      & 1006.5 &  5.0 & 0.41 & 6654 \\
&NLO \S-events (1,1,1) &  991.5 &  6.5 & 0.23 & 3293 \\
&NLO \S-events (2,2,1) &  997.8 &  4.6 & 0.46 & 8522 \\
&NLO \S-events (4,4,1) &  998.7 &  4.1 & 0.59 & 29813 \\
\midrule
\multirow{4}{*}{MC@NLO-$\Delta$}
&RW\_LO (1,1,1)        & 807.8 & 40.4 & 0.004 & 455 \\
&RW\_LO (2,2,1)        & 817.9 & 10.5 & 0.06 & 609 \\
&RW\_LO (4,4,1)        & 792.0 &  7.2 & 0.12 & 1319 \\
&RW\_LO (8,8,1)        & 800.1 &  5.4 & 0.21 & 6623 \\
&RW\_LO (16,16,1)      & 791.3 &  4.8 & 0.28 & 15882 \\
&NLO \S-events (1,1,1) & 787.9 &  6.0 & 0.17 & 6520 \\
&NLO \S-events (2,2,1) & 788.9 &  4.1 & 0.37 & 19402 \\
&NLO \S-events (4,4,1) & 796.5 &  3.6 & 0.50 & 83727 \\
\bottomrule
\end{tabular}
\caption{The \S-event cross section $\sigma$, statistical uncertainty
  $\delta\sigma$ for 100k events, and the Kish effective sample size
  fraction $f_{\mathrm{ESS}}$ for the $pp \to e^+ e^- j$ process, with
  MC@NLO (top rows) and MC@NLO-$\Delta$ (bottom rows) matching.
  The final column shows the generation time of the LHE file.}
\label{tab:stats_zj}
\end{table}

In table~\ref{tab:stats_zj} we present the main results for the $pp
\to e^+ e^- j$ process. We find a very similar pattern as for $pp \to
t\bar{t}$, with the main difference that the fractions of negative
weights, are significantly larger, and therefore the effective sample
sizes $f_{\mathrm{ESS}}$ smaller. This results in larger statistical
uncertainties. In particular, the reweighted sample without any
folding, has such a large spread of event weights that the statistical
power of the sample is so far reduced that effective sample size
fraction is as low as $f_{\mathrm{ESS}}=0.004$ for
MC@NLO-$\Delta$. Even with relatively low levels of folding, the
situation improves considerably, and already with (8,8,1) folding for
the reweighted samples, $f_{\mathrm{ESS}}$ is larger than for the
unweighted NLO \S-events, in a smaller (MC@NLO) or comparable
(MC@NLO-$\Delta$) amount of computation time. With higher levels of
folding, the directly generated NLO \S-events sample outperforms the
reweighted samples in terms of effective sample size, but at a
computational cost. Overall, for a given computational budget, there
is no great difference in $f_{\mathrm{ESS}}$ between generating NLO
\S-events directly, or through reweighting of LO events.

\begin{figure}[!h]
\centering
\includegraphics[width=0.9\linewidth]{"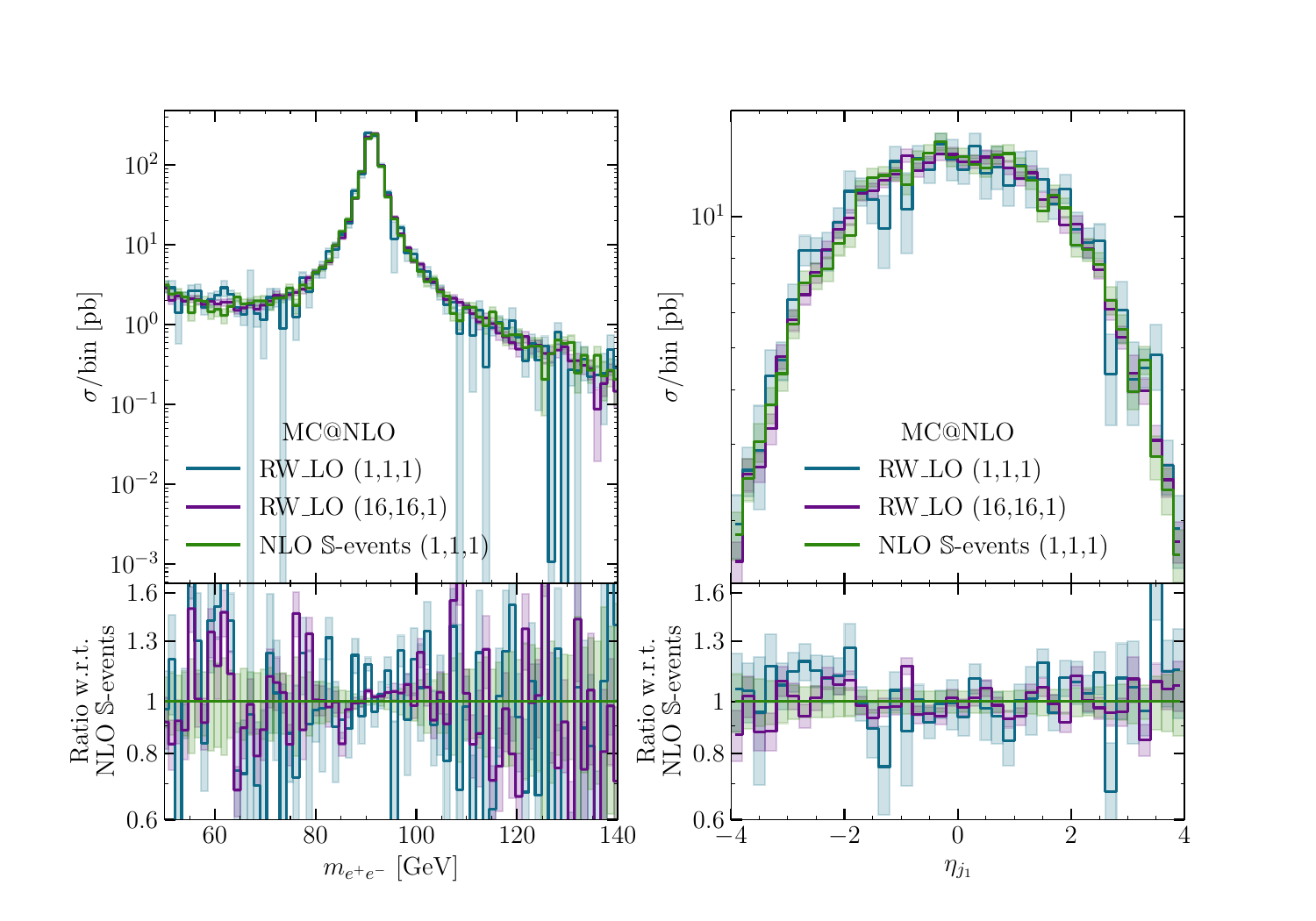"}
\includegraphics[width=0.9\linewidth]{"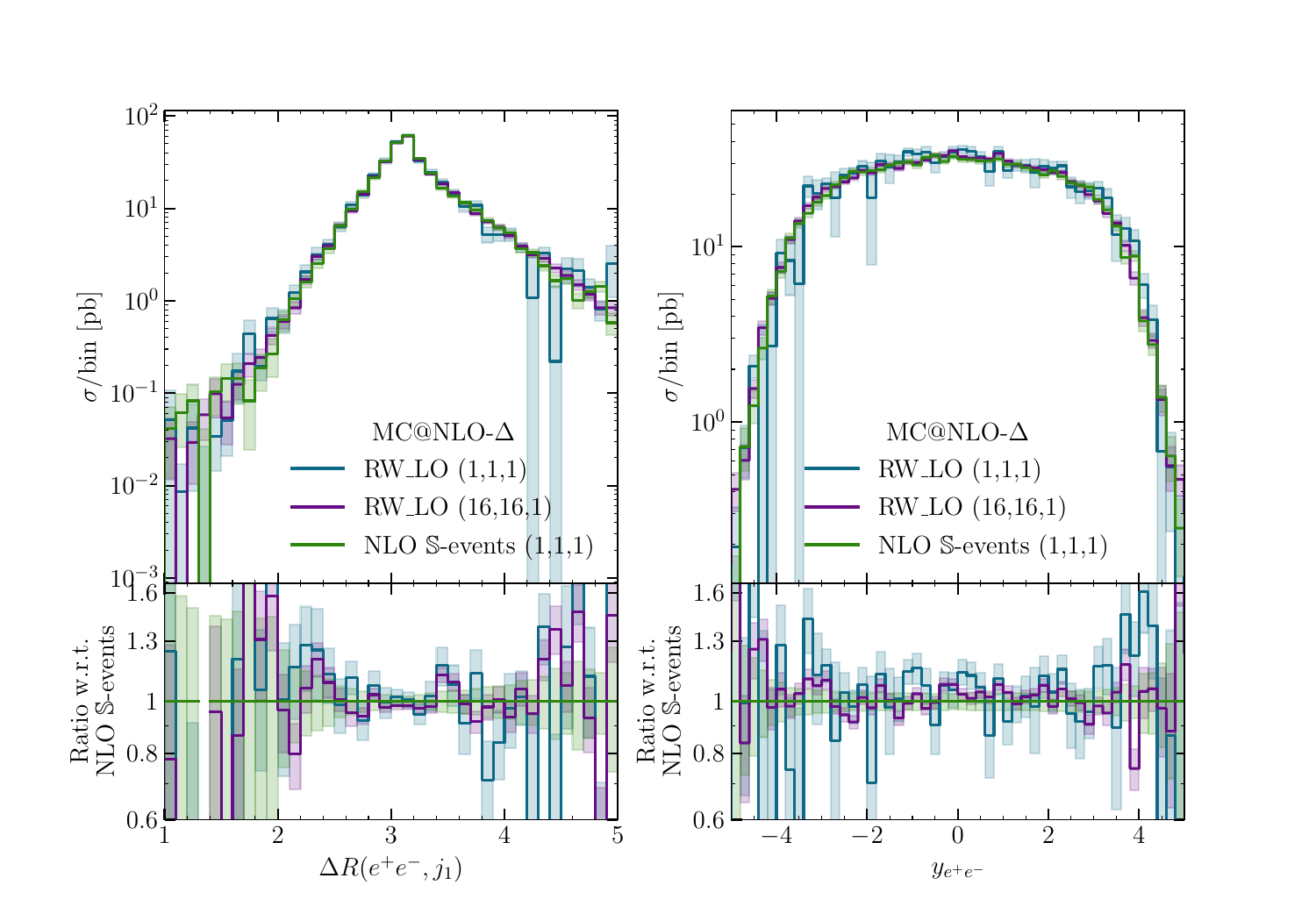"}
\caption{Differential distributions for the \S-event contribution to
  the process $pp \to e^{+}e^{-}j$ comparing reweighted LO samples and
  directly generated NLO S-events.  The top row shows MC@NLO matching:
  the dilepton invariant mass $m_{e^{+}e^{-}}$ (left) and the pseudo-rapidity
  of the hardest jet $\eta_{j_1}$ (right). The bottom row shows the
  MC@NLO-$\Delta$ results: the angular separation between the dilepton
  system and the hardest jet, $\Delta R(e^{+}e^{-},j_1)$ (left), and
  the rapidity of the $e^{+}e^{-}$ pair (right).  Reweighted samples
  without folding suffer from large statistical uncertainties due to
  the broad weight distribution, whereas applying high folding
  $(16,16,1)$ substantially reduces variance and leads to excellent
  agreement with the directly generated NLO S-event predictions.}
\label{fig:zj}
\end{figure}

Also at the differential level, see figure~\ref{fig:zj}, we find the
reweighted samples agree with the directly produced one. The top two
plots are with MC@NLO matching while the bottom two for
MC@NLO-$\Delta$. For MC@NLO, we show the $e^+ e^-$ invariant mass
(left) and the pseudo-rapidity of the hardest jet (right). For
MC@NLO-$\Delta$ we present the separation between the $e^+ e^-$ pair
and the hardest jet in the event, $\Delta R(e^+e^-,j_1)$ (left) and
the rapidity of the $e^+ e^-$ pair (right). For both MC@NLO and
MC@NLO-$\Delta$ the statistical uncertainties are large for the
reweighted samples without folding, but it is greatly reduced with
(16,16,1) folding, outperforming the directly generate \S-events
without folding.

\FloatBarrier
\section{Conclusions}
\label{sec:conclusions}

We have presented and validated an alternative strategy for event
generation at NLO+PS accuracy in the MC@NLO framework, built on
reweighting pre-generated, unweighted Born-level (LO) events to the
full MC@NLO \S-event weight, with \H-events produced in the standard
way. The method keeps the physics content of MC@NLO intact while
decoupling the costly dynamic interplay between $B$, $V$ and the
Monte-Carlo counterterm $K_{\mathrm{MC}}$ from the event-generation
stage. This allows us to invest the saved CPU budget into large
folding factors in the radiative variables, which smooth the integrand
and reduce the variance of \S-event weights.

Across two representative LHC processes, $pp\to t\bar t$ and $pp\to
e^+e^-j$, we find that reweighted \S-event samples reproduce total
rates and differential distributions of a directly generated MC@NLO
reference within statistical uncertainties. At the same time, folding
substantially enhances the Kish effective sample size of the
reweighted samples; for high levels of folding factors it
approaches---and sometimes surpasses---the effective statistics of
unweighted \S-event samples generated without folding. However, the
need for high levels of folding does come at a computational cost. We
find that for $pp\to t\bar t$ production directly generating NLO
samples is more cost-effective than reweighting LO samples, while for
$pp\to e^+e^-j$ the costs are comparable.

Operationally, the implementation is lightweight: it only requires
storing the random numbers (per integration channel) used to generate
each unweighted LO event, and running a dedicated reweighter prior to
file merging. The reweighter computes $V(\Phi_B)$ and the folded
integral over $K_{\mathrm{MC}}$ consistently with the shower profile
that will be used downstream. This minimizes changes to the existing
\mga code base and yields event files that can be showered as usual
(here with \py), ensuring portability and ease of adoption.

The present study suggests several possibilities for optimisation in
the generation of the highly-folded reweighted samples. First, it
would be natural to explore adaptive choices of folding factors that
target regions with large local variance, and to benchmark the
approach for higher-multiplicity and processes with bottom quarks
where negative weights are most severe. Second, a careful assessment
of, for example, adative cubature or Smolyak sparse-grid quadrature
methods for the three FKS radiative variables---possibly with
randomized smoothing to avoid correlated biases---could reduce the
computational costs of the reweighted samples.

In summary, reweighting unweighted LO samples to MC@NLO \S-events,
combined with folding in the radiative variables, provides a viable
alternative route to NLO+PS event generation and merits further
investigation.
\\

\noindent The modified version of \mga, including the reweighter code,
is available from the authors upon request.

\FloatBarrier
\section*{Acknowledgments}
S.E.F.~is supported by the MCnet Short-Term PhD Studentship programme,
funded by CERN, which provided financial support for his research stay.
He gratefully acknowledges the Department of Physics at Lund University
for hosting him during this period and for providing an excellent research
environment.
R.F.~is supported by the Swedish Research Council under contract
number 202004423.

\bibliography{paper}{}
\bibliographystyle{JHEP} 

\end{document}